\documentclass[conference]{IEEEtran}
\IEEEoverridecommandlockouts
\usepackage{cite}
\usepackage{amsmath,amssymb,amsfonts}
\usepackage{graphicx}
\usepackage{textcomp}
\usepackage{xcolor}
\usepackage{comment}
\usepackage{url}
\usepackage{listings}
\usepackage{tabularx}
\usepackage{tikz}
\usepackage{balance}
\usepackage{mciteplus}
\usepackage{subcaption}
\usetikzlibrary{arrows.meta, positioning, shapes.geometric, fit}
\usepackage{hyperref}

\tikzset{
  block/.style={
    draw,
    rounded corners,
    align=center,
    font=\small,
    text width=\linewidth,
    minimum height=1mm,
    inner sep=2.5pt
  },
  line/.style={-Latex, thick}
}

\begin{document}

\title{LatencyLab: A DPDK-Based P4 Pipeline Latency Measurement Framework for FPGA SmartNICs}
\author{
\IEEEauthorblockN{P.~Kuppili\IEEEauthorrefmark{1},
Z.~Han\IEEEauthorrefmark{3},
Y.~Qian\IEEEauthorrefmark{3},
S.~Handagala\IEEEauthorrefmark{3},
M.~Zink\IEEEauthorrefmark{2},
M.~Leeser\IEEEauthorrefmark{3},
R.~Ricci\IEEEauthorrefmark{1}}
\IEEEauthorblockA{\IEEEauthorrefmark{1}University of Utah \quad
\IEEEauthorrefmark{2}University of Massachusetts Amherst \quad
\IEEEauthorrefmark{3}Northeastern University}
}
\maketitle

\begin{abstract}
P4-programmable FPGA SmartNICs place packet processing directly on the wire, but open FPGA P4 toolflows do not expose timestamping at the pipeline boundary, so the latency a P4 program adds on the target FPGA is rarely measured. This paper presents LatencyLab, a DPDK-based measurement framework for FPGA P4 pipeline latency that needs neither PHC/PTP support on the datapath nor clock synchronization. The FPGA's two ports share a network segment, so the switch multicasts a copy of each probe packet to both: one copy passes through the VitisNetP4 pipeline, the other through a matched bypass path. A kernel-bypass DPDK receiver busy-polls both ports and timestamps every packet with the CPU timestamp counter (TSC) as it is retrieved from the NIC's receive circular buffer. The arrival-time difference of the two copies isolates the pipeline latency after calibration against a null bitstream carrying the same traffic; transmit time cancel in the subtraction. We evaluate four VitisNetP4 programs on an AMD Alveo U280, probing each with a 20{,}000-packet trace measured ten times per session over five independent sessions, all TSC-timestamped and reflected for hardware timestamping. The measured latency distributions are tight and reproducible: 99\% of packets fall within 20 ns of the median, session medians repeating within 1--2\,ns (FiveTuple 107/137\,ns, Forward 149\,ns, RemoveHeader 177\,ns, Checksum 364\,ns at 250\,MHz). Two independent checks agree with the framework: a kernel-free reflector returns every probe pair to a ConnectX-5 NIC whose adapter clock reproduces the measured distributions within a few nanoseconds, quantile by quantile, and every measured packet falls 18 to 22 clock cycles below the vendor's worst-case latency bound. The framework also reveals behavior that averages hide: FiveTuple does not have a single latency; at each bring-up it settles into one of two values 31 ns apart, an effect traced to the clock of its lookup memory.

\begin{IEEEkeywords}
Data Plane Development Kit (DPDK), differential measurement, FPGA SmartNICs, hardware timestamping, latency measurement, Open-NIC-Shell, P4, VitisNetP4
\end{IEEEkeywords}

\end{abstract}

\section{Introduction}
\label{sec:intro}
Modern datacenters offload packet processing from hosts into programmable network interface cards (NICs)~\cite{smartnics}, and FPGA SmartNICs programmed in P4 bring that programmability all the way to the wire~\cite{p4vhdl,p4fpga,p4netfpga}. On the open side of this ecosystem, Open-NIC-Shell~\cite{onic} with AMD VitisNetP4~\cite{vitisnetp4} turns an Alveo U280 into a P4-programmable 100\,GbE NIC through a standard toolflow~\cite{p4framework}, and a growing class of applications now runs computation inside that pipeline: sketch-based traffic monitoring maintains count-min counters in the data plane at 100\,Gbps~\cite{sketchfpga}, and stateful network functions built from P4 and high-level synthesis (HLS) extract and aggregate flow information at line rate~\cite{p4hls}. Both systems run on the platform measured in this paper, and both were validated for throughput, but the per-packet latency these pipelines add --- the other half of a real-time processing budget --- could not be measured with existing tools.

That number is rarely measured, because the measurement points do not exist. Direct one-way timing needs timestamps at the pipeline boundary and a synchronized clock, typically a disciplined PTP Hardware Clock (PHC)~\cite{ptp}. Datacenter switches and commodity NICs (cNICs) provide both~\cite{dptp,topocloud}. Open FPGA P4 datapaths provide neither. Previous studies highlight this gap through the compensatory structures they had to construct around it: OP4T needed a custom timestamping overlay because P4-to-FPGA toolflows expose no timestamping primitive at the pipeline boundary~\cite{op4t}, and P4STA inserted a separate stamper FPGA into the path~\cite{p4sta}. Passive FPGA measurement falls back to free-running local counters that no host clock can read~\cite{passivefpga}, and NanoPU reached nanosecond timing only through custom CPU--NIC integration unavailable on off-the-shelf FPGA SmartNICs~\cite{nanopu}. Timing signals are generally unavailable inside FPGA P4 data planes~\cite{ptpinsidep4,onic-linuxptp-issue,ig-tmstmp-vitisnetp4}, and the platforms that do expose PHCs do so through proprietary stacks~\cite{p4ip,iwave} or without open P4 support~\cite{v80}.

The other alternative is to timestamp packets at the destination host, but existing measurement literature cautions against doing so. MoonGen measures latency only with NIC hardware timestamps because CPU-clock host stamps carry PCIe and OS-scheduling variance~\cite{moongen}. \emph{Where Has My Time Gone}~\cite{wheretimegone} and KV-Direct~\cite{kvdirect} show that host and PCIe interactions can outweigh the device latency under study. P4STA~\cite{p4sta} and OSNT~\cite{open-net-test} exist because software timestamping cannot deliver reliable nanosecond accuracy. Our preliminary measurements on this U280/Open-NIC-Shell platform, reported in \ref{subsec:hosteffects}, agree on both counts: return paths that traverse the receiver host bury the pipeline latency under PCIe and kernel effects, which are orders of magnitude larger (Sec.~\ref{subsec:hosteffects}), and kernel receive timestamps carry quantization artifacts of tens of nanoseconds, the same scale as the quantity being measured.

This paper presents \emph{LatencyLab}, a DPDK-based framework that measures the latency distribution of a P4 program on an FPGA-based SmartNIC with packet-level resolution, using only the open toolflow and commodity hardware. Let $L_{p4}$ denote the transit latency of the VitisNetP4 pipeline. LatencyLab estimates $L_{p4}$ differentially. The FPGA's two ports sit on the same network segment, so the switch multicasts a copy of each probe packet to both; one copy is processed by the P4 pipeline while the other crosses a matched bypass path. A kernel-free DPDK receiver busy-polls both ports and timestamps each packet with the CPU timestamp counter (TSC) at the NIC's receive circular buffer, so the per-probe arrival-time difference contains $L_{p4}$ plus a constant path asymmetry. The transmit time and shared network path cancel in the subtraction, and no clocks on different devices need synchronization. The remaining asymmetry is measured using a null bitstream --- the plain shell with no P4 pipeline on either path --- and subtracted from the differential (Sec.~\ref{subsec:protocol}).

Because LatencyLab uses host TSC timestamps rather than NIC hardware timestamps, we verify it against the reference the literature does trust, NIC hardware receive timestamping~\cite{moongen,open-net-test}: every probe pair is reflected, without entering a kernel network stack, to a cNIC that timestamps arrivals in hardware, and its distributions are compared with LatencyLab's. We also check every estimate against the vendor's worst-case latency model~\cite{vitis-latency}. Both agree with the framework (Sec.~\ref{subsec:validation}). The verification is needed once. A deployed LatencyLab requires no PHC, no PTP, and no hardware-timestamping NIC, so it applies where such support is missing.

The contributions of this paper are: 
\begin{itemize}
    \item  LatencyLab, a DPDK-based differential framework for FPGA P4 pipeline latency requiring no datapath timing support (Secs.~\ref{sec:design},~\ref{sec:implementation}).
    \item Characterization of the latency distributions of four VitisNetP4 programs --- reproducible to 1--2\,ns across independent hardware bring-ups (Sec.~\ref{sec:results}).
    \item Verification of the framework against an independent NIC hardware clock and the vendor worst-case model (Sec.~\ref{subsec:validation}).
    \item A case study enabled by LatencyLab’s packet-level resolution: FiveTuple exhibits two observed session-level modes approximately 31 ns apart, consistent with its separately clocked content-addressable memory (CAM) (Sec.~\ref{subsec:ftstates}).
\end{itemize}

\section{Background}
\label{sec:platform}
This section gives the background for LatencyLab: the platform it targets, the P4 programs under test, and the preliminary measurements that informed its design.

\subsection{Testbed}
\label{subsec:testbeds}

\begin{figure*}[t]
    \centering
    \begin{subfigure}[b]{0.49\textwidth}
        \centering
        \includegraphics[width=\linewidth]{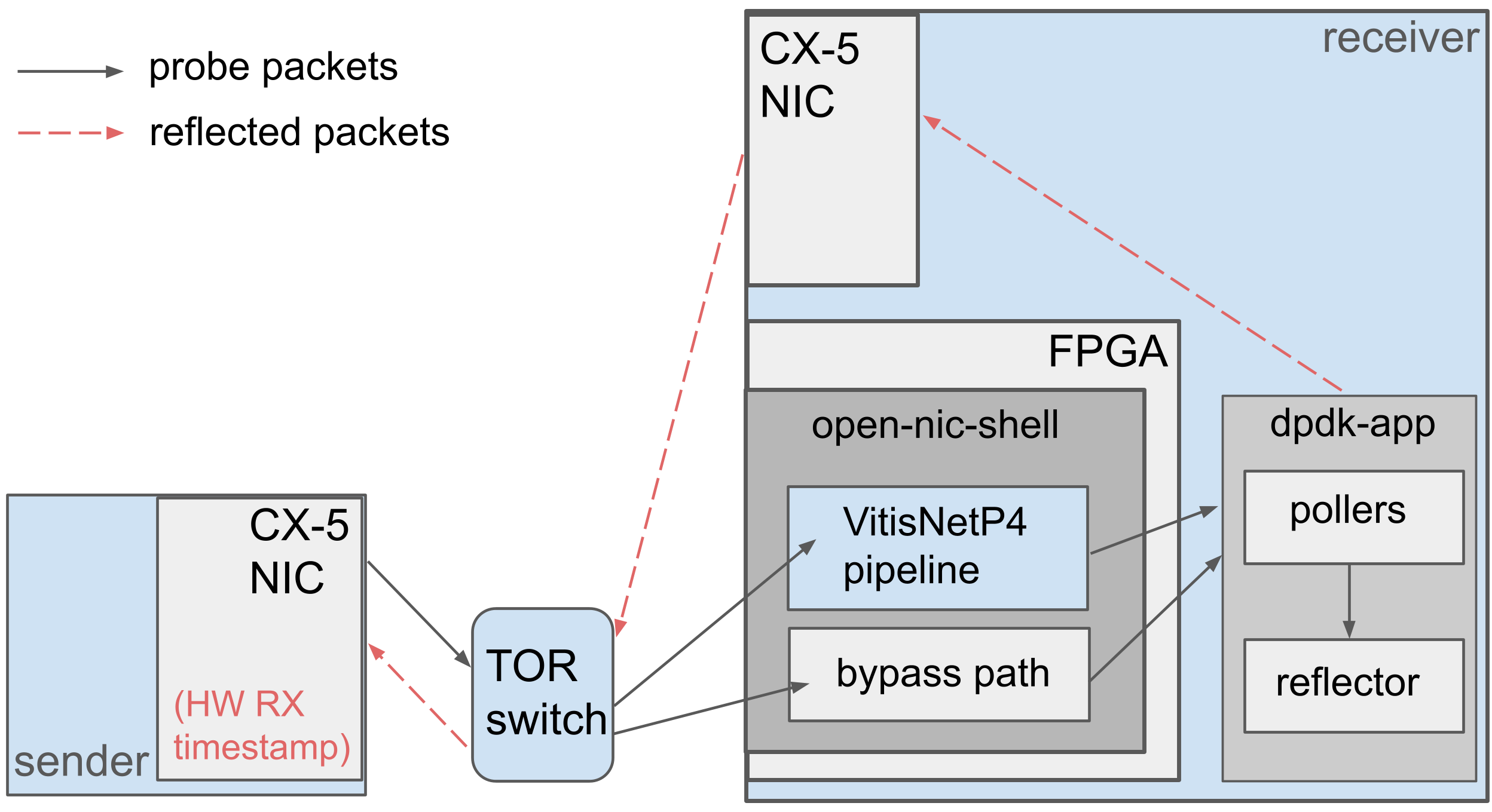}
        \caption{The LatencyLab measurement framework.}
        \label{fig:topology}
    \end{subfigure}
    \hfill
    \begin{subfigure}[b]{0.49\textwidth}
        \centering
        \includegraphics[width=\linewidth]{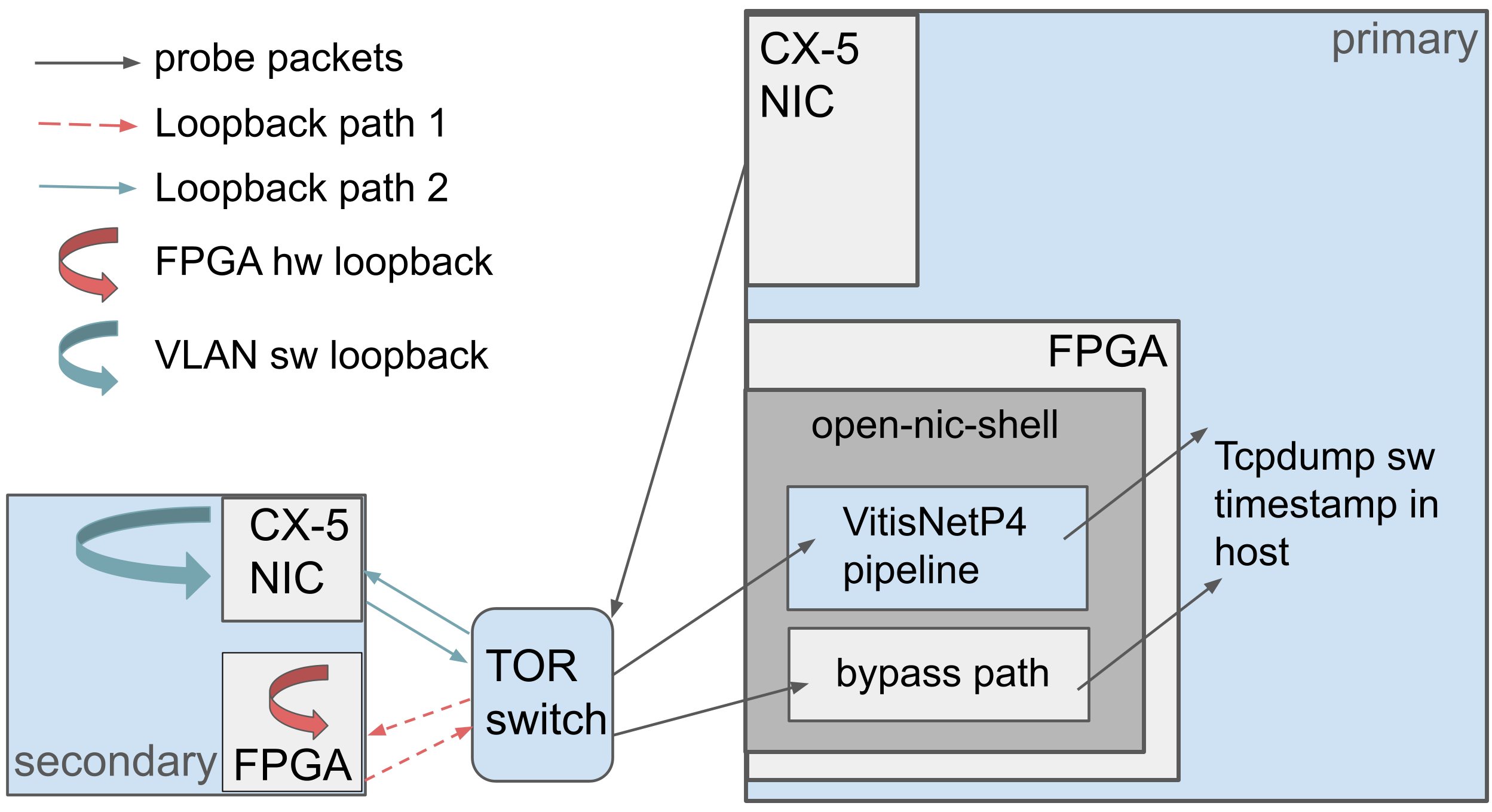}
        \caption{Preliminary host-timed loopback study.}
        \label{fig:hostlooppaths}
    \end{subfigure}
    \caption{Measurement setups on the same OCT slice.}
    \label{fig:setups}
    \vspace{-10pt}
\end{figure*}

All measurements run on the Open Cloud Testbed (OCT)~\cite{oct}, which provides bare-metal servers whose FPGA NIC ports are directly connected to the datacenter switch fabric~\cite{fpga-cloud}. Our slice consists of two servers on a top-of-rack (TOR) switch as shown in Fig.~\ref{fig:topology}. The sender node has one 100\,GbE NVIDIA ConnectX-5 cNIC connected to the TOR. The receiver node hosts an AMD Alveo U280~\cite{fpga-cloud} with two 100\,GbE connections directly to the TOR switch using QSFP28 transceivers and one 100\,GbE NVIDIA ConnectX-5 cNIC. The sender replays probe traffic from its cNIC; the receiver hosts the FPGA under test.

The receiver uses a dual-socket Intel Xeon Gold 6226R (2.90 GHz, 16 cores per socket) host running Ubuntu 22.04 LTS (kernel 5.15.0-177); we use Open-NIC-Shell (commit a33fb0b)~\cite{onic}, open-nic-dpdk with Xilinx dma\_ip\_drivers (commit 7859957)~\cite{onic-dpdk}, DPDK 20.11, VitisNetP4 2023.2, and Vivado 2023.2. Both DPDK polling cores are pinned to cores 4 and 6 on the same processor socket (NUMA node 0).

\subsection{Why Open-NIC-Shell}
\label{subsec:whyonic}

\begin{figure}[ht]
    \centering
    \includegraphics[width=\linewidth]{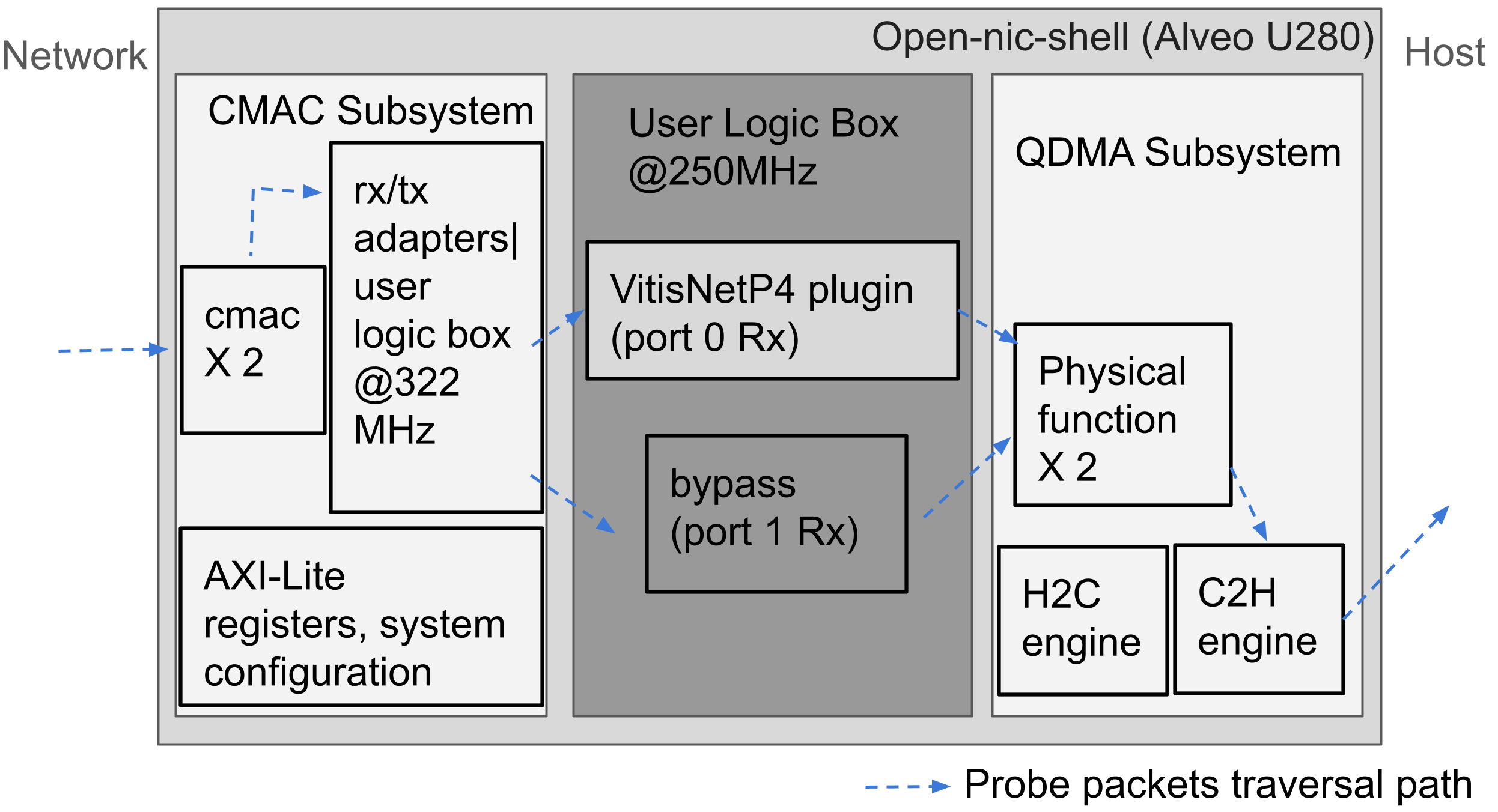}
    \caption{Open-NIC-Shell datapaths as used by LatencyLab.}
    \label{fig:onicarch}
    \vspace{-10pt}
\end{figure}

We build on Open-NIC-Shell~\cite{onic} for two reasons. First, it is the de facto open-source NIC shell for AMD FPGAs and the standard path for deploying P4 on shared research testbeds such as OCT~\cite{p4framework}, which is where the pipeline-cost question arises in practice. Second, and essential to the framework, its QDMA subsystem has an open DPDK poll-mode driver, which allows LatencyLab to timestamp packets as they are retrieved from the NIC's receive circular buffer with no kernel involvement (Sec.~\ref{subsec:instrument}).

As shown in Fig.~\ref{fig:onicarch}, Open-NIC-Shell exposes the two 100\,GbE ports of the FPGA as two PCIe physical-function interfaces on the host. One port's receive path carries the VitisNetP4 IP (interface $PF_{\mathrm{vitis}}$); the other carries a passthrough plugin ($PF_{\mathrm{bypass}}$). Everything else about the two paths is common: the same 250\,MHz user-logic clock, the same QDMA subsystem, the same PCIe Gen3 x16 link. For measurement purposes this is the property that matters: the processing paths are architecturally identical apart from the VitisNetP4 pipeline itself. Consequently, any components common to both paths effectively cancel out in a differential comparison.

\subsection{P4 Programs Under Test}
\label{subsubsec:p4bit}

We evaluate four programs derived from the AMD VitisNetP4 example designs~\cite{amd-examples}, built with Vivado~\cite{vivado}. \emph{FiveTuple} extracts the 5-tuple and matches it in a content-addressable memory (CAM) before forwarding; its lookup engine is generated with a 300\,MHz CAM clock alongside the 250\,MHz packet-pipeline clock, a detail that matters in Sec.~\ref{subsec:ftstates}. \emph{Forward} parses Ethernet/VLAN/IPv4/IPv6/TCP/UDP and forwards on an IPv4 destination match. \emph{RemoveHeader} strips the VLAN header and forwards the shortened packet. \emph{Checksum} parses the same headers as \emph{Forward} and additionally computes and matches the IPv4 header checksum in the data plane; it is the deepest pipeline of the four. A fifth bitstream, the \emph{null}, is the plain shell with passthrough/bypass plugins on \emph{both} ports and no P4 pipeline anywhere; it serves as the calibration reference in Sec.~\ref{subsec:protocol}.

\subsection{What Host-Path Timing Shows: PCIe and Kernel Dominate}
\label{subsec:hosteffects}

\begin{figure}[ht]
    \centering
   \includegraphics[width=\columnwidth]{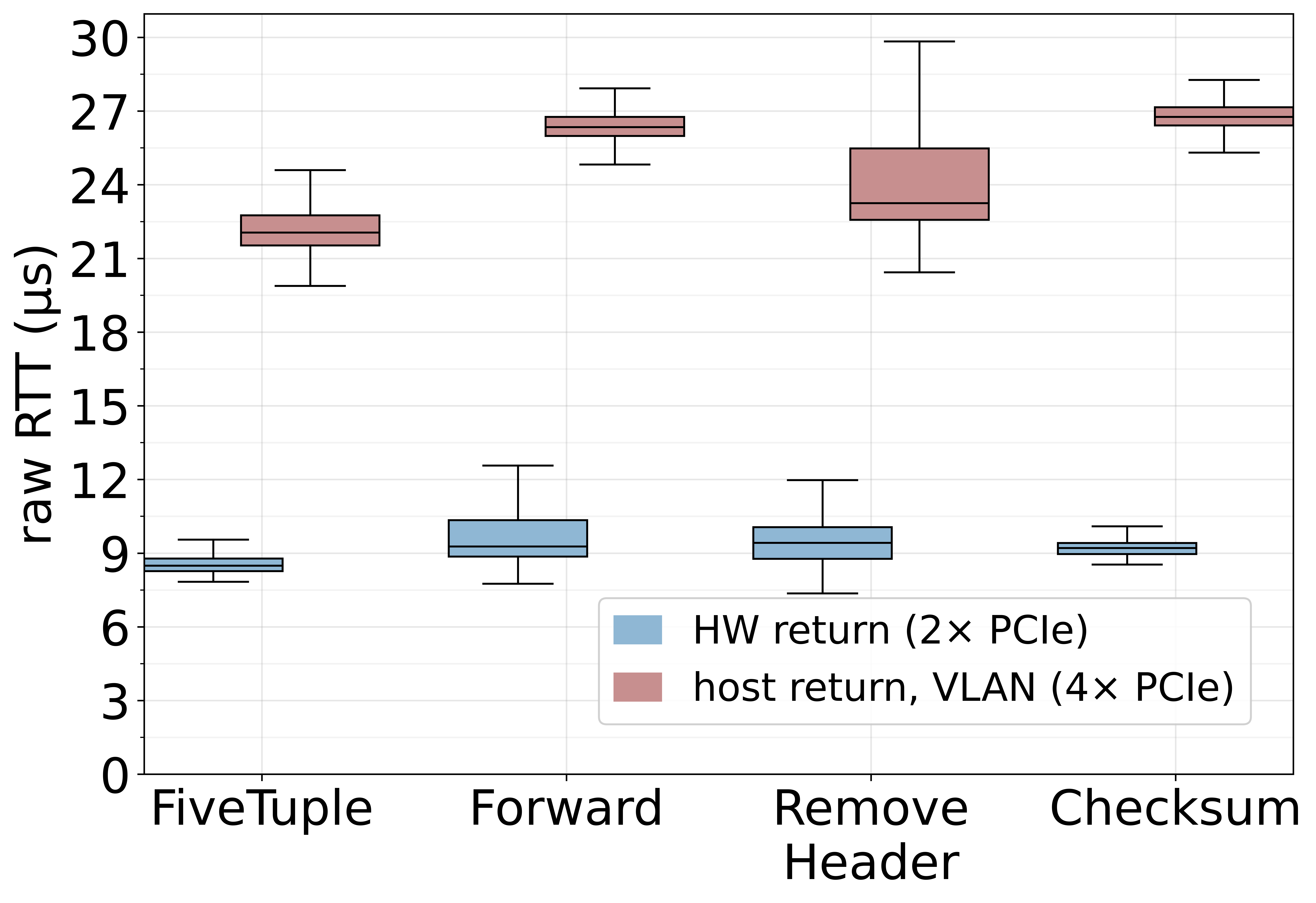}
    \caption{Raw RTT of the two loopback paths; the sub-$\mu$s pipeline latency is invisible in both.}
    \label{fig:hosteffects}
    \vspace{-10pt}
\end{figure}

Before building the framework, we measured what host-visible timing can actually resolve, using an earlier arrangement of the same testbed in which the primary node hosts both the cNIC and the FPGA, as shown in Fig.~\ref{fig:hostlooppaths}. Probe packets are replayed with \texttt{tcpreplay} from the primary node's cNIC and looped back to the same node's FPGA ports, where tcpdump records software timestamps in the host; the loop is closed in one of two ways. In the first, a secondary node connected over the TOR switch reflects each packet in its AMD Alveo U280 in hardware, so the packet never touches a host and the round trip includes two PCIe crossings, both on the primary node. In the second, the secondary node reflects packets in software through a VLAN bridge, which adds another traversal of its PCIe link and kernel network stack, for a total of four PCIe crossings. Fig.~\ref{fig:hosteffects} compares the raw RTT distributions of the two paths. The result confirms the findings in~\cite{wheretimegone,kvdirect}: both RTT distributions operate at the scale of microseconds, the two added host crossings visibly inflate both the median and the spread, and the approximately 0.1 - 0.4 µs pipeline contribution is unresolved within either distribution. This result motivates the work presented in this paper.

\section{The LatencyLab Framework}
\label{sec:design}
Sec.~\ref{subsec:hosteffects} leaves two requirements: cancel everything the two paths share, and timestamp packets before the kernel can distort them. LatencyLab meets the first with a differential design, where the switch delivers a copy of each probe to both FPGA ports and the paired arrival times are subtracted. The second is addressed with a DPDK receiver that timestamps every packet with the CPU timestamp counter (TSC). One term survives the subtraction, the constant asymmetry between the two ports themselves; a null bitstream with no P4 pipeline on either path measures it directly, so it is measured and removed rather than assumed negligible.

\subsection{Overview}
\label{subsec:latencylaboverview}

Fig.~\ref{fig:topology} shows the measurement framework design. The sender replays a trace of probe packets, each carrying a unique ID in its payload. Both FPGA ports are connected to the same network segment, so the switch multicasts a copy of each probe to each FPGA port. One copy is processed by the VitisNetP4 pipeline on its way to $PF_{\mathrm{vitis}}$; the other crosses the passthrough plugin to $PF_{\mathrm{bypass}}$. A DPDK application on the receiver polls both ports and timestamps every arriving packet, and the ID ties the two copies of a probe back together. The copies share the transmit and upstream network path but enter distinct FPGA ports and QDMA queues, their arrival-time difference removes the shared terms and retains the pipeline latency plus a constant port-and-queue asymmetry. The transmit time and the shared path drop out in the subtraction, which is why no clocks on different devices need to be synchronized. The rest of this section fills in each stage.

\subsection{Sender and Probe Traffic}
\label{subsec:sender}

Probe traffic for each program is taken directly from its AMD VitisNetP4 example design~\cite{amd-examples} and built into a 20{,}000-packet trace, with a unique ID written into every payload. The sender replays the trace with \texttt{tcpreplay} at 500 packets per second, one probe every 2\,ms, and this rate is a deliberate choice rather than a limitation. The quantity we want is transit latency. At any rate where probes can queue behind each other, in the switch, in the pipeline, in the DMA engines, or at the poller, the measurement would also include queueing delay, which depends on load rather than on the program. Spacing probes 2\,ms apart makes each probe an isolated event through the entire system. The slow rate also protects timestamp quality on the receiver: each copy arrives in its own DMA burst, so the poller handles it alone and stamps it immediately, in the same way on both ports. Higher offered loads would characterize platform queueing rather than isolated pipeline transit latency; we leave that distinct question to future work.

\subsection{Kernel-Free Receiver}
\label{subsec:instrument}

The receiver is a DPDK application on Open-NIC-Shell's~\cite{onic-dpdk} QDMA poll-mode driver, with the FPGA's two functions bound to \texttt{vfio-pci}. One dedicated core polls each port in a busy loop; no interrupt, kernel stack, or system call touches the packet path. For each received packet the poller reads the TSC using \texttt{rte\_rdtsc\_precise()} immediately after \texttt{rte\_eth\_rx\_burst()} dequeues the packet from the NIC receive ring, and then parses the embedded ID. Both pollers run on cores of the same processor socket and read hardware-synchronized invariant TSCs, so their timestamps require no cross-device clock conversion. TSC cycles are converted to nanoseconds using the TSC frequency reported by DPDK's \texttt{rte\_get\_tsc\_hz()},(2.89 GHz on this host); the CPU's \texttt{constant\_tsc/nonstop\_tsc} capability guarantees the TSC ticks at this fixed rate regardless of core frequency scaling. Any fixed intercore offset is included in the calibrated asymmetry \(\delta\).

\subsection{Reflect Framework for Verification}
\label{subsec:witnessdesign}

Since LatencyLab’s TSC timestamps constitute a custom-designed timing mechanism, we validate them against NIC hardware receive timestamps~\cite{moongen,open-net-test}, which are used by prior measurement systems as a reference for nanosecond-scale latency. Any NIC that can hardware-timestamp arbitrary received frames can serve as this reference; we used the ConnectX-5 already present in the sender. The challenge is delivering the packets to that NIC without them passing through any kernel network stack, since that would reintroduce the very timing noise of Sec.~\ref{subsec:hosteffects}; the reflect path shown in Fig.~\ref{fig:topology} solves this. Immediately after a packet is timestamped, the same DPDK process returns it through the receiver's cNIC (attached through the bifurcated mlx5 driver) toward the sender's cNIC, which timestamps every arriving frame against its own adapter clock (\texttt{HWTSTAMP\_FILTER\_ALL} with nanosecond precision). The reflect path must preserve two things about each pair: which copy is which, and their timing comparability. A one-byte tag in the payload handles the first, marking each frame as the VitisNetP4 or the bypass copy, since both arrive at the same NIC; a fixed TX queue per path handles the second, turning any queue-to-queue difference into a constant per-path offset that the analysis absorbs, rather than per-packet noise that would blur the distributions being compared. The adapter-clock arrival difference of a reflected pair re-measures the differential measured by LatencyLab, plus noise from the reflection channel itself; Sec.~\ref{subsec:validationmodel} turns this into a precise consistency test. The reflect framework exists only for verification. A deployed LatencyLab does not require this component, nor any hardware-timestamping NIC.

\subsection{Null Calibration}
\label{subsec:protocol}

One term survives the subtraction: the two ports are distinct physical ports with distinct QDMA queues, so their delivery paths differ by a small constant. Modeling this asymmetry would mean trusting assumptions about port internals, so LatencyLab measures it instead. The null bitstream (Sec.~\ref{subsubsec:p4bit}) makes both paths passthrough while changing nothing else. Replaying the same probe trace using the null bitstream therefore yields the port-and-queue asymmetry for that trace, whose median is subtracted from the program differential. The measured offsets are small and stable, 14 to 28\,ns depending on the trace, with a session-to-session spread under 7\,ns. Without this step, every reported latency would carry an unknown bias of that magnitude.

\section{Measurement Model}
\label{sec:implementation}
This section defines LatencyLab's measurement protocol and the quantity it computes.
Measurements are organized into sessions. A session is one complete hardware instantiation (bitstream flash, PCIe re-enumeration, driver and table initialization, DPDK start) followed by ten replay runs, and we collect five independent sessions per program, with each session observing $PF_{\mathrm{vitis}}$ and $PF_{\mathrm{bypass}}$ concurrently. Each instantiation puts the hardware into a fresh state: a property that is set at reset and then held until the next reset stays constant within a session but can differ across sessions. Collecting five independent sessions therefore separates reset-time effects from per-packet behavior, a choice that is justified in Sec.~\ref{subsec:ftstates}. In total, each program is measured with 20,000 probes $\times$ 10 runs $\times$ 5 sessions per program, with both paths observing the same probes. Every probe is TSC-timestamped on both ports and reflected for hardware timestamping, so both clocks observe the same packets. We repeat the same session protocol with the null bitstream for calibration.

\subsection{Differential Estimator}
\label{subsec:p4latencymeasurement}

Consider probe $i$, transmitted at time $T(i)$, whose two copies arrive at the receiver. Writing $S(i)$ for the delay shared by both copies (wire, switch, and everything else upstream of the split), the two TSC arrival stamps are:

\begin{align}
t_P(i) &= T(i) + S(i) + L_{p4}(i) + d_P + e_P(i), \label{eq:1}\\
t_B(i) &= T(i) + S(i) + d_B + e_B(i), \label{eq:2}
\end{align}

\noindent
where $d_P, d_B$ are the constant per-path delivery components (port, queue, DMA engine) and $e_P, e_B$ run-varying measurement error caused by polling phase and PCIe/DMA timing. Their difference,

\begin{equation}
D(i) = t_P(i) - t_B(i) = L_{p4}(i) + \delta + \varepsilon(i), \label{eq:3}
\end{equation}

\noindent
retains only the pipeline latency, the constant asymmetry $\delta = d_P - d_B$, and noise $\varepsilon = e_P - e_B$. The transmit time and the shared path components are eliminated.

Eq.~\eqref{eq:3} still contains the port asymmetry $\delta$, and the null bitstream supplies the equation that removes it. With no P4 pipeline on either path, the same measurement gives:

\begin{equation}
D_0(i) = \delta + \varepsilon_0(i). \label{eq:4}
\end{equation}

\noindent
Subtracting Eq.~\eqref{eq:4} from Eq.~\eqref{eq:3} cancels $\delta$ and leaves $L_{p4}$ plus noise. In practice the subtraction uses $\hat{\delta}$, the median null-bitstream differential for the corresponding traffic trace $D_0$, and the measurement error is mitigated through aggregation across ten runs per session: the latency of probe $i$ repeats in every run while the noise does not, so the median preserves the recurring packet-dependent component while reducing the influence of run-specific timing variation. We use the median rather than the mean because one atypical polling, DMA, or PCIe delay can substantially shift the mean, whereas the median of ten repeats is robust to such timing outliers. The estimate is:

\begin{equation}
\hat{L}_{p4}(i) = \operatorname*{med}_{r=1..10} D_r(i) \;-\; \hat{\delta}. \label{eq:5}
\end{equation}

\noindent
Sec.~\ref{sec:results} reports, for each program, the distribution of $\hat{L}_{p4}(i)$ over all probes and sessions.

\subsection{Verifying the Framework}
\label{subsec:validationmodel}

We perform two independent evaluations to evaluate the differential estimator in Eq.~\eqref{eq:5} and present the results in Sec.~\ref{subsec:validation}.

\subsubsection{Hardware-clock check} The reflect framework (Sec.~\ref{subsec:witnessdesign}) re-times each probe pair in the sender cNIC's adapter clock, observing $R(i) = D(i) + C(i)$ where $C(i)$ is reflection-channel noise generated by devices and clocks disjoint from LatencyLab. If LatencyLab measures $D$ correctly and $C$ is independent of it, the distribution of $R$ must equal the convolution of the measured $D$ distribution with that of the paired residuals $C(i)=R(i)-D(i)$; randomly re-pairing residuals with differentials synthesizes this convolution, and comparing its quantiles against the directly measured $R$ tests the framework at packet-pair granularity with no shared clock or timestamping mechanism.

\subsubsection{Vendor-model check} VitisNetP4 reports a worst-case \emph{calculated latency} for each generated pipeline, extended by a traffic-dependent penalty only when the program inserts headers~\cite{vitisnetp4}; none of our four programs does, so the bound is the calculated latency itself: 53, 55, 66, and 109 cycles at 250\,MHz for FiveTuple, Forward, RemoveHeader, and Checksum, respectively. Since this constitutes an upper bound rather than a deterministic prediction, consistency with the model requires every retained $\hat{L}_{p4}(i)$ to remain below the corresponding threshold, when the margin exhibits uniform behavior across programs --- consistent with conservative bound properties --- independent of program scale.


\section{Results and Analysis}
\label{sec:results}
\subsection{Latency of the Four Programs}
\label{subsec:mainresult}
\label{subsec:validation}

\begin{figure}[t]
    \centering
    \includegraphics[width=\linewidth]{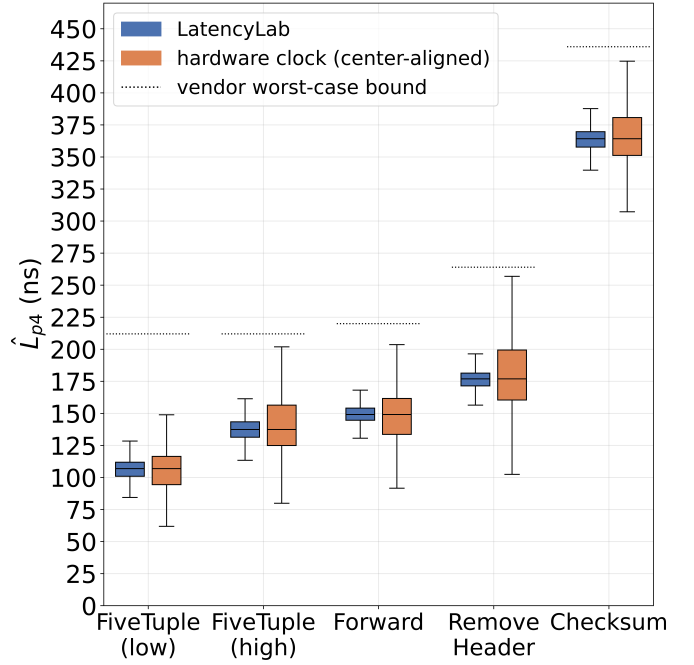}
    \caption{Pipeline latency of the four programs.}
    \label{fig:main}
    \vspace{-10pt}
\end{figure}

\begin{table*}[t]
    \centering
    \caption{Per-program P4 pipeline latency}
    \footnotesize
    \setlength{\tabcolsep}{4pt}
    \renewcommand{\arraystretch}{1.15}
    \begin{tabular*}{\textwidth}{@{\extracolsep{\fill}}|l|cccccc|cccccc|cc|}\hline
        & \multicolumn{6}{c|}{\textbf{LatencyLab (ns)}} & \multicolumn{6}{c|}{\textbf{Hardware validation (ns)}} & \multicolumn{2}{c|}{\textbf{Vendor bound}} \\
        \textbf{P4 Program} & \textbf{p5} & \textbf{p25} & \textbf{p50} & \textbf{p75} & \textbf{p95} & \textbf{p99}
        & \textbf{p5} & \textbf{p25} & \textbf{p50} & \textbf{p75} & \textbf{p95} & \textbf{p99}
        & \textbf{(ns)} & \textbf{(cycles)} \\\hline
        FiveTuple (low)  &  92 & 101 & 107 & 112 & 120 & 127 &  75 &  94 & 107 & 116 & 133 & 151 & 212 & 53 \\\hline
        FiveTuple (high) & 124 & 131 & 137 & 143 & 152 & 159 & 111 & 125 & 137 & 156 & 182 & 204 & 212 & 53 \\\hline
        Forward          & 138 & 145 & 149 & 154 & 163 & 169 & 114 & 134 & 149 & 162 & 185 & 207 & 220 & 55 \\\hline
        RemoveHeader     & 157 & 171 & 177 & 181 & 188 & 192 & 128 & 160 & 177 & 199 & 224 & 241 & 264 & 66 \\\hline
        Checksum         & 349 & 358 & 364 & 370 & 377 & 381 & 332 & 351 & 364 & 381 & 399 & 412 & 436 & 109 \\\hline
    \end{tabular*}
    \label{table:results}
    \vspace{-10pt}
\end{table*}

Fig.~\ref{fig:main} and Table~\ref{table:results} show the main results of the paper. For each program, the figure places three things side by side: the latency measured by LatencyLab, the same probes re-measured by the independent hardware clock, and the vendor worst-case bound. All numbers are per-probe estimates: each probe is measured ten times, and Eq.~\eqref{eq:5} takes the median to remove single-measurement noise. The three sources tell one consistent story. The LatencyLab distributions are narrow (p5--p99 spans of 31--36\,ns, p99 within 22\,ns of the median), which is what a feed-forward pipeline should produce; the residual spread is remaining measurement noise plus genuine parsing differences across packets in a trace. The hardware clock, which sees the same differential through the noisier reflection channel, results in very similar distributions but with a wider percentile span. Every distribution sits well below its vendor bound. The programs separate cleanly and order by pipeline depth, from FiveTuple (26.7 cycles in its low state) up to Checksum (91 cycles). The results are reproducible: across five full hardware re-initializations per program, raw session medians stay within a 1--2\,ns band (165--167\,ns for Forward, 386--390\,ns for Checksum). 

\begin{figure}[t]
    \centering
    \includegraphics[width=0.88\linewidth]{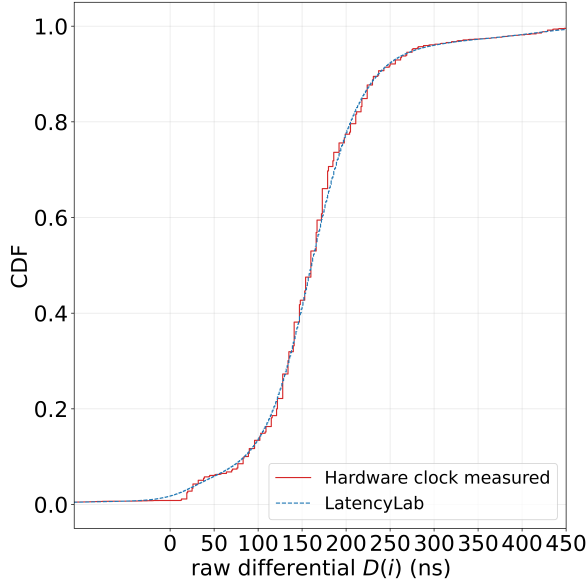}
    \caption{Measured hardware-clock distribution vs.\ the convolution predicted from LatencyLab's measurements (Forward, Sec.~\ref{subsec:validationmodel}).}
    \label{fig:conv}
    \vspace{-10pt}
\end{figure}

The agreement between the two clocks in Fig.~\ref{fig:main} is quantitative, not just visual. Fig.~\ref{fig:conv} shows the convolution check of Sec.~\ref{subsec:validationmodel} for Forward: the measured and predicted distributions agree within 1\,ns at p50 and p90 and within 4\,ns at p25--p99, and across all four programs the agreement holds within 8\,ns at every quantile from p25 through p95 (the HW column of Table~\ref{table:results}). Where the reflection channel preserves packet size (Forward), the centers also agree, to 9\,ns; programs that alter packet length (RemoveHeader, FiveTuple) shift the reflected copies' serialization by a program-dependent constant, which is why the hardware-clock results in Fig.~\ref{fig:main} and Table~\ref{table:results} are center-aligned. The hardware clock therefore validates the distribution shape for all programs, and the absolute center when the reflection path preserves packet size.

Every per-probe estimate lies below its program's bound, with one systematic exception: the first probe after an idle gap arrives late by tens of nanoseconds. This is not a pipeline effect. It affects only packet one of a run, appears identically for every program regardless of pipeline depth, and the feed-forward pipeline holds no state that could warm up; the receive path performs its first DMA activity after the inter-run idle interval. After excluding these startup probes, all retained estimates remain below their bounds, and even p99 leaves 12--18 cycles of margin. The headroom at the median is nearly constant, 18.7, 17.7, 21.8, and 18.0 cycles for FiveTuple’s high mode, Forward, RemoveHeader, and Checksum, across pipelines whose depths differ by $2.5\times$. One program does not have a single latency, and we examine it next.

\subsection{Special Case: FiveTuple's Two Latency States}
\label{subsec:ftstates}

\begin{figure}[t]
    \centering
    \includegraphics[width=\linewidth]{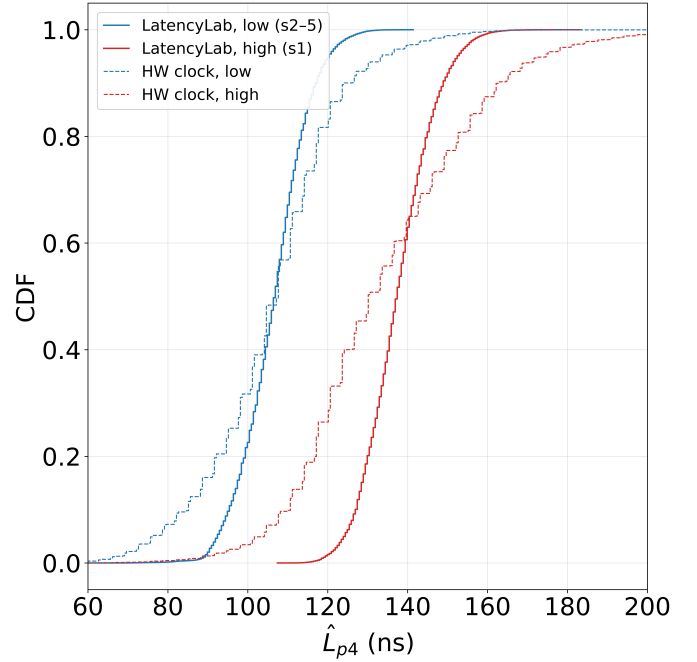}
    \caption{FiveTuple raw session medians as measured by LatencyLab (right) and, by the hardware clock of the reflect framework (left).}
    \label{fig:ftstates}
    \vspace{-10pt}
\end{figure}

FiveTuple's five sessions do not result in similar latencies. Four sessions (2--5) sit at 105--107\,ns while session 1 sits at 137\,ns, a step of about 31\,ns (7.7 cycles) with no intermediate values and no within-session transition across the ten runs, as shown in Fig.~\ref{fig:ftstates}. The hardware clock of the reflect framework saw the same split from entirely different hardware on the same day: with the reflection channel's constant removed (Sec.~\ref{subsec:validation}), its low sessions sit at 104--110\,ns and session 1 at 126\,ns. The state is real, and it is chosen at instantiation: the latency is decided once, when the FPGA comes up, and does not change again until the next flash and reset.

The likely reason is the CAM's separate clock. FiveTuple is the only program of the four whose lookup engine runs on its own clock: VitisNetP4 instantiates the CAM at 300\,MHz alongside the 250\,MHz packet-pipeline clock~\cite{vitisnetp4}. A signal crossing between two clock domains costs a number of cycles that depends on how the two clocks' edges align~\cite{cdc}, and on this platform both clocks derive from the same on-chip clock generator, so their alignment is set when the FPGA comes up and holds until the next reset. This mechanism predicts the qualitative behavior we observe: a latency chosen per bring-up, quantized in cycles, constant within a session, and absent from the three single-clock programs. The P4 control flow also applies the table from two call sites, so the crossing penalty is paid twice per packet.

This case study shows what the framework buys. A host-timed measurement would drown a 31\,ns step in PCIe and kernel spread (Sec.~\ref{subsec:hosteffects}). A single-session measurement could observe only one mode and miss the reset dependence. A measurement averaged across sessions could report a value between the observed modes that the hardware never exhibits. Resolving the behavior took packet-level differentials, and a session-structured protocol. The practical takeaway is that a FiveTuple-class pipeline has at least two stable per-reset latencies in the measured sessions, which matters for latency budgeting.


\section{Conclusion}
\label{sec:conclusion}
This paper presented LatencyLab, a DPDK-based framework that measures the latency a P4 program adds on an FPGA SmartNIC, with packet-level resolution and without the timing support such platforms lack: no datapath PHC, no PTP synchronization between devices, and no kernel in the timestamp path. The design uses what the platform already provides. The shell's two host-visible ports receive copies of the same probe, which turns the problem into an arrival-time differential in which everything shared cancels. The shell's DPDK support lets one process timestamp both ports at the NIC's receive circular buffer with a single invariant clock. A null bitstream measures and removes the constant port-and-queue offset that the differential cannot cancel.

On an AMD Alveo U280, we evaluate four VitisNetP4 programs using 20,000 probes, each measured ten times per session across five independent sessions. For each reported program or FiveTuple latency mode, the 99th-percentile latency is at most 22\,ns above the median—meaning that 99\% of the per-probe latency estimates are no greater than the median plus 22\,ns. Session medians are reproducible to 1--2,ns across bring-ups for the single-clock programs, while FiveTuple exhibits two reset-dependent modes. Two independent checks agree with the framework: a NIC hardware clock re-timed every probe pair through the kernel-free reflect framework and reproduced the measured distributions, and the vendor's worst-case model bounds every estimate with similar headroom across all four programs. The FiveTuple case study shows what this resolution is worth. The program has two latency modes, 31\,ns apart, consistent with the relative alignment of its CAM and packet-pipeline clock domains.

Future work includes characterizing queueing onset at high offered load, probing the FiveTuple state ladder across many resets, and porting to other Open-NIC-Shell-class boards. Our bitstreams, measurement applications, and dataset are publicly available in the LatencyLab GitHub repository at \url{https://github.com/OCT-FPGA/LatencyLab}.

\section{Acknowledgement}
\label{sec:acknowledgement}
This work was funded by National Science Foundation (NSF) grants 2319962, 2130907, 2130891, 2027208, and 2431419. All opinions and statements are those of the authors and do not represent the position of the NSF. The authors thank the Open Cloud Testbed team for maintaining the FPGA infrastructure on which all measurements were performed.


\bibliographystyle{IEEEtran}
\bibliography{references}

\end{document}